\documentclass[12pt,epsfig]{article}
\usepackage{graphicx}

\usepackage[latin1]{inputenc}
\usepackage{amsmath}
\usepackage{amsfonts}
\usepackage{amssymb}
\usepackage{graphicx}
\DeclareGraphicsExtensions{.pdf,.png,.jpg}

\usepackage{epstopdf}
\def\beq{\begin{equation}}
\def\eeq{\end{equation}}
\def\bea{\begin{eqnarray}}
\def\eea{\end{eqnarray}}

\begin{document}

\begin{center}
	{\Large \bf New scattering features in non-Hermitian space fractional quantum mechanics
	}

\vspace{1.3cm}

{\sf   Mohammad Hasan  \footnote{e-mail address: \ \ mhasan@isac.gov.in, \ \ mohammadhasan786@gmail.com}$^{,3}$,
 Bhabani Prasad Mandal \footnote{e-mail address:
\ \ bhabani.mandal@gmail.com, \ \ bhabani@bhu.ac.in  }}

\bigskip

{\em $^{1}$ISRO Satellite Centre (ISAC),
Bangalore-560017, INDIA \\
$^{2,3}$Department of Physics, Institute of Science,
Banaras Hindu University,
Varanasi-221005, INDIA. \\ }

\bigskip
\bigskip

\noindent {\bf Abstract}

\end{center}
The spectral singularity (SS) and coherent perfect absorption (CPA) have been extensively studied over the last one and half decade for different non-Hermitian potentials in non-Hermitian standard quantum mechanics (SQM) governed by Schrodinger equation. In the present work we explore these scattering features in the domain of non-Hermitian space fractional quantum mechanics (SFQM) governed by fractional Schrodinger equation  which is characterized by Levy index $\alpha$ ($ 1< \alpha \leq 2$). We observe  that  non-Hermitian SFQM systems have more flexibility for SS and CPA and display some new features of scattering. For the delta potential $V(x)=-i\rho \delta (x-x_{0})$, $\rho > 0$, the SS energy, $E_{ss}$, is blue or red shifted with decreasing $\alpha$ depending the strength of the potential. For complex rectangular barrier in non-Hermitian SQM, it is known that the reflection and transmission amplitudes are oscillatory near the spectral singular point. It is found that these oscillations eventually develop SS in non-Hermitian SFQM. The similar features is also reported for the case of CPA phenomena from complex rectangular barrier in non-Hermitian SFQM. These observations suggest a deeper relation between scattering features of non-Hermitian SQM and  non-Hermitian SFQM.   

\medskip
\vspace{1in}
\newpage

\section{Introduction}
Certain class of non-Hermitian systems with real energy eigenvalues have become the topic of 
frontier research over the last two decades because one can have fully consistent quantum theory
by restoring the equivalent Hermiticity and upholding the unitary time evolution for such system in a
modified Hilbert space \cite{ben4}-\cite{benr}. Some of the predictions of non-Hermitian quantum theory have been verified experimentally in optics \cite{ opt1}-\cite{eqv1}. The experimental realization of non-Hermitian systems in laboratories prompted huge interests in the study of non-Hermitian systems both in theory and in experiments \cite{km}-\cite{hsn}.  The scattering features of non-Hermitian Hamiltonian have many novel properties which are absent in the Hermitian systems. The scattering features  
due to non-Hermitian potential like exceptional points (EPs) \cite{ep0}-\cite{ep2}, spectral singularity 
(SS) \cite{ss1}-\cite{ss3}, invisibility  \cite{aop}-\cite{inv1}, reciprocity  \cite{aop}-\cite{resc}, coherent perfect absorption (CPA) \cite{cpa00}-\cite{cpa4}
 and critical coupling (CC) \cite{cc0}-\cite{cc4} have generated huge interests during last few 
years due to their applicability and usefulness in the study of different optical systems. The CPA which is actually the time-reversed counterpart of a laser has become the center to all such studies in optics due to the discovery of anti-laser in which incoming beams of light interfere with one another in such a way as to perfectly cancel each
other out.

Soon after the Hermitian quantum mechanics was fruitfully extended to non-Hermitian Hamiltonian, a new generalization of quantum mechanics in its path integral approach was also introduced. This new generalization of standard qunatum mechanics has been termed as fractional quantum mechanics. The concept of fractals  in quantum mechanics was first introduced in the context of path integral (PI) formulation of quantum mechanics \cite{feynman}. In the PI formulation, the path integrals are taken over Brownian paths which results in Schrodinger equation of motion. Nick Laskin generalized the path integral approach of quantum mechanics by considering the path integrals over Levy flight paths \cite{1,2}. The Levy flight paths are characterized by a parameter $ \alpha $ known as Levy index .  This natural generalization of quantum mechanics is termed as space fractional quantum mechanics (SFQM) and is governed by fractional Schrodinger equation also known as Laskin equation. The range of $\alpha$ in SFQM is limited to $1< \alpha \leq 2$ \cite{2}. As for $\alpha=2$ the Levy flight paths are the Brownian paths, this ensures that all the results of  standard quantum mechanics are the special case  of SFQM with Levy index $\alpha=2$.  The field of SFQM has grown fast over the last one and half decade and various application of SFQM have been discussed in quantum mechanics \cite{3,4,5,6}, in optics  \cite{longhi_fractional_optics} and in the context of tunneling time \cite{tt_sfqm}.  
\paragraph{}  
The tunneling problem of SFQM have been solved for various potential configuration by many authors \cite{4,8,tare}. However to the best of our knowledge the study of SFQM has not been extended for non-Hermitian potential which accounts for gain and loss configuration. In the present work we study the scattering feature SS and CPA in the domain of SFQM for complex delta and rectangular barrier potential. For case of complex delta potential we analytically find the SS energy and show that depending upon the strength of the delta potential, the SS energy is either blue are red shifted with decreasing $\alpha$. The SS and CPA features are numerically investigated for complex rectangular barrier potential. It is found that with decreasing $\alpha$ both SS and CPA energy are blue shifted. This shows that the non-Hermitian SFQM system have more flexibility to obtain SS or CPA at desired energy provided the parameter $\alpha$ is controllable. We also found that for complex rectangular barrier potential, the oscillations in reflection ($R$) and transmission amplitude ($T$) around SS energy for $\alpha=2$ slowly grows as  $\alpha$ reduces and eventually develop SS for some value of $\alpha <2$. The similar phenomena is also seen for the case of CPA. This is a new features of scattering and shows that the oscillations in $R$ and $T$ around SS (or CPA) in non-Hermitian SQM has a direct relation with development of SS (or CPA) in non-Hermitian SFQM. For complex rectangular barrier case we only discuss numerically the behaviour of SS and CPA with varying $\alpha$. A more detailed theoretical work towards the realization of such system will be attempted in our future work.

We organize our paper as follows. In section \ref{fse} we discuss the fractional Schrodinger equation. The scattering features of complex delta potential in SFQM are discussed in section \ref{sfqm_complex_delta}. In section \ref{sfqm_complex_barrier} we discuss the SS and CPA for complex barrier potential in SFQM. Final section \ref{the_result} is kept for the conclusions and discussions. 
 
\paragraph{}

\section{The fractional Schrodinger equation}
\label{fse}
The fractional Schrodinger equation in one dimension is
\begin{equation}
i \hbar \frac{\partial \psi (x,t)}{\partial x}= H_{\alpha} (x,t) \psi(x,t)  \ \ \ \ , 1<\alpha \leq2
\label{tdfse}
\end{equation}
where $H_{\alpha} (x,t)$ is the fractional Hamiltonian operator and is expressed through Riesz fractional derivative $(-\hbar^{2} \Delta)^{\alpha/2}$ as
\begin{equation}
H_{\alpha} (x,t)=D_{\alpha} (-\hbar^{2} \Delta)^{\frac{\alpha}{2}} +V(x,t)
\end{equation}
$D_{\alpha}$ is a constant which depends on the system and $\Delta=\frac{\partial^{2}}{\partial x^{2}}$. The Riesz fractional derivative of the wave function $\psi(x,t)$ is defined through its Fourier transform $\tilde{\psi}(p,t)$ as
\begin{equation}
(-\hbar^{2}\Delta)^{\frac{\alpha}{2}} \psi(x,t)=\frac{1}{2\pi \hbar} \int_{-\infty} ^{\infty} { \tilde{\psi}(p,t)\vert p \vert ^{\alpha} e^{\frac{ipx}{\hbar}}dp } 
\label{Riesz_fractional_derivative}
\end{equation}
The Fourier transform of $\psi(x,t)$ is
\begin{equation}
\tilde{\psi}(p,t)= \int_{-\infty} ^{\infty} \psi(x,t) e^{-i\frac{px}{\hbar}} dx
\end{equation}
and
\begin{equation}
\psi(x,t)=\frac{1}{2\pi \hbar} \int_{-\infty} ^{\infty} \tilde{\psi}(p,t) e^{i\frac{px}{\hbar}} dp
\end{equation}
when potential $V(x,t)$ is independent of time we have the time independent fractional Hamiltonian operator $H_{\alpha}(x)=D_{\alpha} (-\hbar^{2} \Delta)^{\frac{\alpha}{2}} +V(x)$. In this case the time independent fractional Schrodinger equation is
\begin{equation}
D_{\alpha} (-\hbar^{2} \Delta)^{\frac{\alpha}{2}}\psi(x)+V(x)\psi(x)=E\psi(x)
\label{tifse}
\end{equation} 
where $E$ is the energy of the particle and $\psi(x,t)=\psi(x)e^{-\frac{iEt}{\hbar}}$. We will be using this formulation to discuss the scattering features of non-Hermitian potentials in the following sections.

\section{Scattering features of complex delta potential in SFQM}
\label{sfqm_complex_delta}
\begin{figure}
\begin{center}
\includegraphics[scale=0.6]{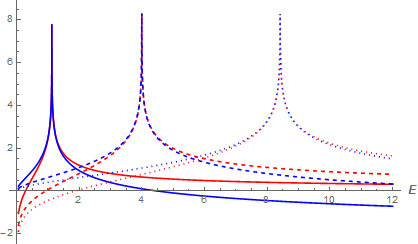}    
\caption{\it The spectral singularity for the delta potential $V(x)=-1.5i \delta (x-x_{0})$ for different values of $\alpha$. Here  $v=10^{-5}$. The red and blue curve represent the variations of $\log_{10} R_{l}$ and $\log_{10} T_{l}$ respectively with energy $E$. From left to right, the continuous, dashed and dotted curve are for $\alpha$ values $2$, $1.99$ and $1.85$ having SS energy at $1.125$, $3.995$ and $8.409$ respectively. The SS energy is blue shifted as $\alpha$ decreases when $2 \hbar v < \rho$.} 
\label{1gt2gt3}
\end{center}
\end{figure}

Consider the complex delta potential
\beq
V(x)=\zeta \delta(x-x _{0}) 
\label{delta_potential}
\eeq
where height of the potential $\zeta$ is a complex number. Following \cite{tare} the transfer matrix 
\beq
 M= \begin{pmatrix}   m_{11} & m_{12} \\ m_{21} & m_{22}   \end{pmatrix}  
\eeq
for this potential in space fractional quantum mechanics has following elements:
The diagonal elements,
\begin{eqnarray}
m_{11}  &= &1+i\zeta (2D_{\alpha} k_{\alpha}^{\alpha -1} \hbar^{\alpha})^{-1}   \label{m11_delta} \\
m_{22} &=&1-i\zeta (2D_{\alpha} k_{\alpha}^{\alpha -1} \hbar^{\alpha})^{-1} \label{m22_delta}
\end{eqnarray}
and the off-diagonal elements
\begin{eqnarray}
m_{12} &=& i\zeta (2D_{\alpha} k_{\alpha}^{\alpha -1} \hbar^{\alpha})^{-1}    \label{m12_delta} \\
m_{21}  &=&-i\zeta (2D_{\alpha} k_{\alpha}^{\alpha -1} \hbar^{\alpha})^{-1} \label{m21_delta}
\end{eqnarray}
with
\begin{equation}
k_{\alpha}=\left( \frac{E}{D_{\alpha} \hbar^{\alpha}} \right)^{\frac{1}{\alpha}}
\label{kalpha}
\end{equation}
We take the generalized diffusion coefficient $D_{\alpha}$ as \cite{tare} 
\begin{equation}
D_{\alpha}  =\frac{v^{2-\alpha}}{\alpha m^{\alpha-1}}
\label{dalpha}
\end{equation}
where $v$ is the characteristic velocity of the non-relativistic system. At $E=E_{ss}$, $E_{ss} \in R^{+}$ one has the spectral singularity if $m_{22}(E_{ss})=0$ \cite{ss1}. Therefore we have to solve the following equation for real positive $E$ such that
\begin{equation}
1-i\zeta (2D_{\alpha} k_{\alpha}^{\alpha -1} \hbar^{\alpha})^{-1}=0
\label{delta_ss_equation}
\end{equation} 
We write $\zeta= \vert \zeta \vert e^{i \phi}$ and with the aid of Eq. \ref{kalpha}, the solution of Eq. \ref{delta_ss_equation} is
\begin{equation}
E_{ss}=\left( \frac{1}{D_{\alpha} \hbar^{\alpha}} \right)^{\frac{1}{\alpha -1}} \left( \frac{\vert \zeta \vert}{2}\right)^{\frac{\alpha}{\alpha -1}} e^{\frac{i \alpha}{\alpha -1} \left(\frac{\pi}{2} +\phi \right)}
\label{e_ss_delta}
\end{equation}
It is evident from Eq. \ref{e_ss_delta} that for $E_{ss}$ to be real, we must have $\phi=-\frac{\pi}{2}$. Hence $\zeta=-i \rho$, $\rho \in R^{+}$. Thus the delta potential $V(x)=-i\rho \delta(x-x_{0})$, ($\rho \in R^{+}$) admits spectral singularity at the energy
\begin{equation}
E_{ss}=\left( \frac{1}{D_{\alpha} \hbar^{\alpha}} \right)^{\frac{1}{\alpha -1}} \left( \frac{ \rho }{2}\right)^{\frac{\alpha}{\alpha -1}} 
\label{ess_delta}
\end{equation}
In terms of the characteristics velocity $v$ of the non relativistic system
\begin{equation}
E_{ss}=m \ v^{\frac{\alpha -2}{\alpha -1}}\left( \frac{\alpha}{\hbar^{\alpha}}\right)^{\frac{1}{\alpha -1}}  \left(\frac{\rho}{2} \right)^{\frac{\alpha}{\alpha -1}}
\label{ess_delta_v}
\end{equation}
Eq. \ref{ess_delta_v} shows that with increasing $v$, $E_{ss}$ reduces for a given $\alpha$. For a given $\alpha$, $E_{ss}$ increases with $\rho$. 
Let $E_{ss}^{\alpha_{1}}$ and $E_{ss}^{\alpha_{2}}$ be the SS energy for $\alpha$ values  $\alpha_{1}$ and $\alpha_{2}$ respectively then for fix $\rho$ and $v$ it can be shown that
\begin{equation}
\frac{E_{ss}^{\alpha_{1}}}{E_{ss}^{\alpha_{2}}}= \left(\frac{\alpha_{1}^{\frac{1}{\alpha_{1}-1}}}{\alpha_{2}^{\frac{1}{\alpha_{2}-1}}} \right) \left( \frac{2 \hbar v}{\rho}\right)^{\frac{\alpha_{1}-\alpha_{2}}{(\alpha_{1}-1)(\alpha_{2}-1)}}
\end{equation}
The purpose of above derivation is to show how $E_{ss}$ varies with $\alpha$. For $\alpha \rightarrow 1$, $\alpha^{\frac{1}{\alpha -1}} \rightarrow e$ where $e$ is the exponential constant.  For $\alpha=2$, $\alpha^{\frac{1}{\alpha -1}}$ is $2$.  Therefore for $\alpha_{1} < \alpha_{2}$, the first term in parenthesis of right hand side is greater than unity. Therefore if  
\begin{equation}
\frac{2 \hbar v}{\rho} < 1
\label{inequality_ess_greater} 
\end{equation}
then  $E_{ss}^{\alpha_{1}} > E_{ss}^{\alpha_{2}}$. This is graphically shown in Fig \ref{1gt2gt3}. All plots are obtained in units $\hbar=1$, $m=1$, $c=1$.
 
Substituting $\alpha_{2}=2$ and $\alpha_{1}=\alpha$ 
\begin{equation}
\frac{E_{ss}^{\alpha}}{E_{ss}}= \left(\frac{\alpha^{\frac{1}{\alpha-1}}}{2} \right) \left( \frac{2 \hbar v}{\rho}\right)^{\frac{\alpha-2}{(\alpha-1)}}
\label{ess_alpha1_standard}
\end{equation}
\begin{figure}
\begin{center}
\includegraphics[scale=0.7]{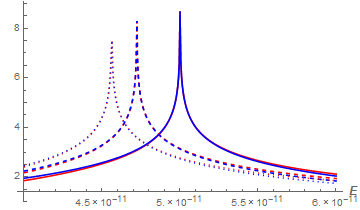}    
\caption{\it The spectral singularity for the delta potential $V(x)=-10^{-5}i \delta (x-x_{0})$ for different values of $\alpha$. Here  $v=10^{-5}$. The red and blue curve represent the variations of $\log_{10} R_{l}$ and $\log_{10} T_{l}$ respectively with energy $E$. From right to left, the continuous, dashed and dotted curve are for $\alpha$ values $2$, $1.99$ and $1.85$ having SS energy at $5 \times 10^{-11}$, $4.72 \times 10^{-11}$ and $4.56 \times 10^{-11}$ respectively. The SS energy is red shifted as $\alpha$ decreases when $2 \hbar v > e \rho/2 $.} 
\label{red_shift_figure}
\end{center}
\end{figure}

where $E_{ss}$ is the SS energy in non-Hermitian SQM. The first term in parenthesis of right hand side is greater than unity for the range $1< \alpha <2$. Therefore if the condition \ref{inequality_ess_greater} is satisfied  then we always have $E_{ss} \leq E_{ss}^{\alpha} < \infty$ for $1< \alpha \leq 2$. Thus we have shown that if $E_{ss}$ is the SS energy of the complex delta potential $V(x)=-i\rho \delta(x-x_{0})$, ($\rho \in R^{+}$) in non-Hermitian SQM and if the condition \ref{inequality_ess_greater} is also fulfilled then one can have SS at any energy $E> E_{ss}$ in non-Hermitian SFQM with a suitably chosen Levy index $\alpha$. Since the maximum value of $\alpha^{\frac{1}{\alpha-1}}/2 $ is $\frac{e}{2}$ for $1< \alpha \leq 2$ therefore for the following case\begin{equation}
\frac{2 \hbar v}{\rho} > \frac{e}{2}
\label{inequality_ess_lesser}
\end{equation}    
we always have $0 < E_{ss}^{\alpha} \leq E_{ss}$ for $1 <\alpha \leq 2$, i.e. with decreasing $\alpha$ the SS energy is red shifted. This case is graphically shown in Fig \ref{red_shift_figure}. For the case $2 \hbar v=\rho$ we have $ E_{ss} \leq E_{ss}^{\alpha} < \frac{e}{2} E_{ss}$ for   $1 < \alpha \leq 2$.

\section{Scattering features of complex barrier potential in SFQM}
\label{sfqm_complex_barrier}
Consider the non-Hermitian barrier potential  $V(x)=V=V_{1}+iV_{2}$, $\{V_1,V_2\} \in R$ over the interval $(0,b)$ and zero elsewhere. The transfer matrix for this potential in SFQM has the following elements \cite{tare}
\begin{eqnarray}
m_{11}  &= &  (\cos{ \overline{k}_{\alpha}b} -i\mu_{1}\sin{ \overline{k}_{\alpha}b})e^{ik_{\alpha}b}  \label{m11_barrier} \\
m_{22} &=& (\cos{ \overline{k}_{\alpha}b} +i\mu_{1}\sin{ \overline{k}_{\alpha}b})e^{-ik_{\alpha}b} \label{m22_barrier}
\end{eqnarray}
and the off-diagonal elements
\begin{eqnarray}
m_{12} &=& i \mu_{2} \sin{ \overline{k}_{\alpha}b}    \label{m12_barrier} \\
m_{21}  &=&-i \mu_{2} \sin{ \overline{k}_{\alpha}b} \label{m21_barrier}
\end{eqnarray}
where,
\begin{eqnarray}
\mu_{1}=\frac{1}{2}\left(\varepsilon+\frac{1}{\varepsilon} \right)  \label{mu1_formula}  \\
\mu_{2}=\frac{1}{2}\left(\varepsilon-\frac{1}{\varepsilon} \right) \label{mu2_formula} 
\end{eqnarray} 
\begin{equation}
\varepsilon=\left( \frac{k_{\alpha}}{\overline{k}_{\alpha}}\right)^{\alpha-1}
\label{epsilon_formula}
\end{equation} 
and
\begin{equation}
\overline{k}_{\alpha}=\left( \frac{E-V}{D_{\alpha} \hbar^{\alpha}}\right)^{\frac{1}{\alpha}}
\label{kalphabar}
\end{equation}
where $k_\alpha $ and $ D_{\alpha}$ are given by Eqs. \ref{kalpha}, \ref{dalpha} respectively. From the elements of transfer matrix, the reflection and transmission coefficients are obtained as
\begin{equation}
t_{l}=\frac{1}{m_{22}} \ \ , r_{l}=\frac{m_{21}}{m_{22}}
\label{tl_rl}
\end{equation} 

\begin{equation}
t_{r}=\frac{1}{m_{22}} \ \ , r_{r}=\frac{m_{12}}{m_{22}}
\label{tr_rr}
\end{equation} 
The corresponding amplitudes are $T_{l,r}=|t_{l}=t_{r}|^{2}$ , $R_{l,r}=|r_{l,r}|^{2}$.  The spectral singularity corresponds to the simultaneous blow up of reflection and transmission amplitude at a particular energy  for unidirectional incidence. At $E=E_{ss}$, $E_{ss} \in R^{+}$, one has the spectral singularity if $m_{22}(E_{ss})=0$ \cite{ss1} i.e. the real energy at which the zeros of $m_{22}$ occur.  The condition for coherent perfect absorption is 
\begin{equation}
t_{l} (E)t_{r}(E)- r_{l}(E)r_{r}(E)=0
\label{cpa_condition1}
\end{equation}
for $E=E_{CPA} \in R^{+}$. We discuss SS and CPA in the subsequent section for complex barrier potential in SFQM. 
\subsection{Spectral singularity in non-Hermitian SFQM}
\begin{figure}
\begin{center}
\includegraphics[scale=0.5]{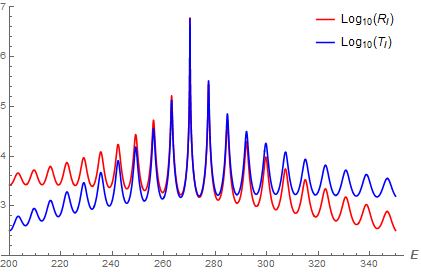} a \includegraphics[scale=0.5]{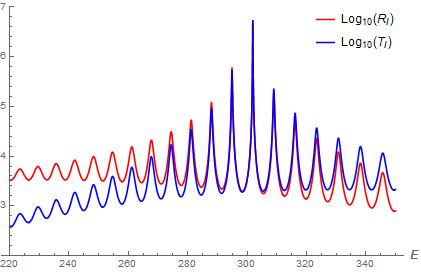} b \\
\includegraphics[scale=0.5]{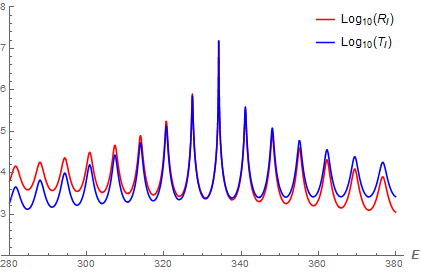} c \includegraphics[scale=0.5]{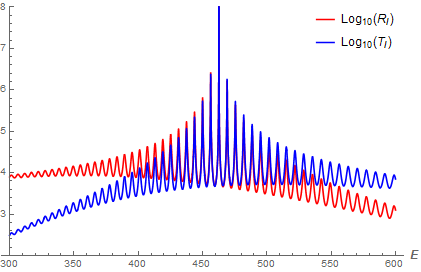} d   \\
\caption{\it The spectral singularity for the potential $V=9.1675-10i$ for different values of $\alpha$. Here $b=10$ and $v=10^{-5} $.The red and blue curves represent $\log_{10}{R_{l}}$ and $\log_{10}{T_{l}}$ respectively. For the figure $a$, $b$, $c$ and $d$ the values of $\alpha$ are $2$, $1.99$, $1.98$ and $1.95$ respectively. The SS energy is blue shifted as $\alpha$ decreases.} 
\label{fig_ss}
\end{center}
\end{figure}

\begin{figure}
\begin{center}
\includegraphics[scale=0.5]{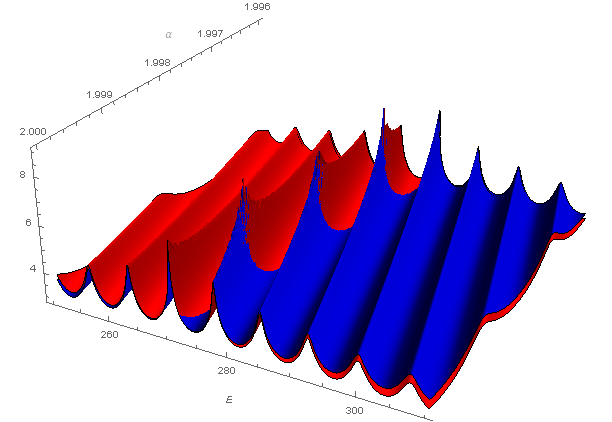} a  \\
\includegraphics[scale=0.5]{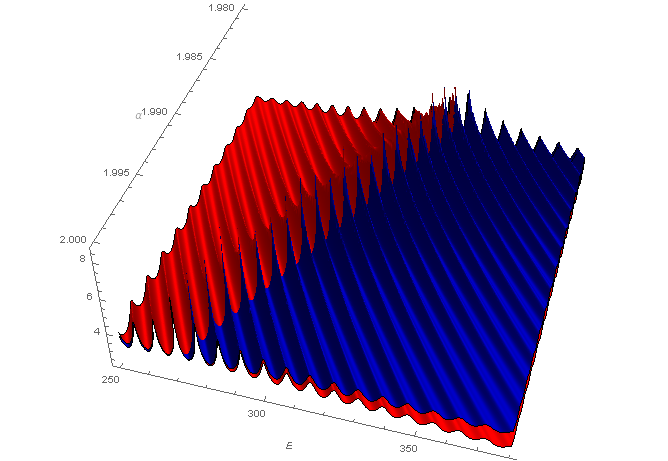} b   \\
\caption{\it The development of multiple spectral singularities for $\alpha \leq 2$ in $\alpha -E$ space. Red and blue plots represent $\log_{10}{R}$ and $\log_{10}{T}$ respectively. The potential parameters are the same as of Fig \ref{fig_ss}. The range of $\alpha$ in Fig-a above is $1.996 \leq \alpha \leq 2$ and in Fig-b it is $1.98 \leq \alpha \leq 2$. It is observed that peak of the oscillations in $\log_{10}{R}$ , $\log_{10}{T}$ for $\alpha =2$ and $E> E_{ss}^{\alpha=2}$ slowly amplify with decreasing $\alpha$ and develop SS for some values of $\alpha<2$. $E_{ss}^{\alpha=2}$ is the SS energy in non-Hermitian SQM.} 
\label{fig_ss_suboscillations}
\end{center}
\end{figure}

\begin{figure}
\begin{center}
\includegraphics[scale=0.6]{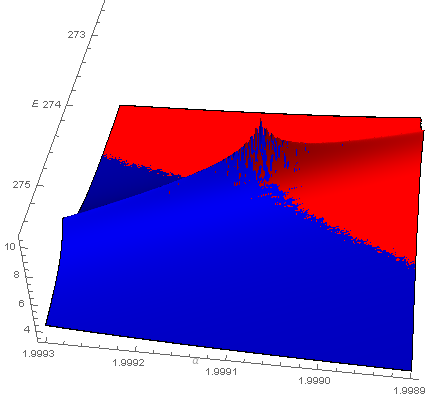} 
\caption{\it Close view of the development of SS over $\alpha -E$ space for the first sub-peak of Fig \ref{fig_ss_suboscillations}. Red and blue plots represent $\log_{10}{R}$ and $\log_{10}{T}$ respectively. The potential parameters are the same as of Fig \ref{fig_ss} (or Fig \ref{fig_ss_suboscillations}). This plot shows that in the neighbourhood of SS there are oscillations in $R$ and $T$ with $\alpha$ .} 
\label{fig_ss_closelook}
\end{center}
\end{figure}

\begin{figure}
\begin{center}
\includegraphics[scale=0.7]{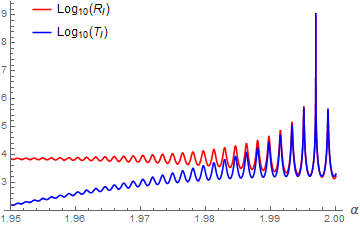}  
\caption{ \it Graphical illustrations of sub-peaks around SS with $\alpha$ for $E=280$. Again the potential parameters are the same as of Fig \ref{fig_ss}.} 
\label{ss_suboscillation_alpha}
\end{center}
\end{figure}
As mentioned before at $E=E_{ss}$, $E_{ss} \in R^{+}$, one has the spectral singularity if $m_{22}(E_{ss})=0$. This corresponds to solve the following equation 
\begin{equation}
\cos{ \overline{k}_{\alpha}b} +i\mu_{1}\sin{ \overline{k}_{\alpha}b}=0
\end{equation}
for $E \in R^{+}$.
This is a transcendental equation and has to be solved numerically. For the case of non-Hermitian SQM  this transcendental equation has been solved for  non-Hermitian potential barrier in an essentially analytical method \cite{ss4}. In the present work we numerically find spectral singularity. We study the variation of $R_{l}$ and $T_{l}$ (we will drop the subscript $l$) with energy $E$ of the particle to obtain SS. For complex barrier potential, SS  for different values of $\alpha$ is shown in the Fig. \ref{fig_ss}. From the figure it is evident that as $\alpha$ decreases, the divergence in $R$ and $T$ is blue shifted. To understand the development of these divergence  for $\alpha<2$  we ($3$D) plot $\log_{10}R$ and $\log_{10}T$ with $E$ and $\alpha$. This $3$D plot is shown in Fig \ref{fig_ss_suboscillations} for the same potential parameter as in Fig \ref{fig_ss}. In Fig \ref{fig_ss_suboscillations} -a, the range of $\alpha$ is $1.996 \leq \alpha \leq 2$. We observe that the simultaneous peak (and the maxima) in $\log_{10}R$ and $\log_{10}T$ for $\alpha=2$ occur around $E=E_{ss}^{\alpha=2}=270.11$.  To the both sides of this energy there are peaks in $\log_{10}R$ and $\log_{10}T$ due to their oscillations with E for $E>E_{ss}^{\alpha=2}$ and  $E<E_{ss}^{\alpha=2}$. We will call these peaks in $\log_{10}R$ and $\log_{10}T$ around $E_{ss}^{\alpha=2}$ as `{\it sub-peaks}' around $E_{ss}^{\alpha=2}$. From Fig \ref{fig_ss_suboscillations}-a one observe that the sub-peaks for $E>E_{ss}^{\alpha=2}$ grow in amplitude as $\alpha$ decreases till the simultaneous maxima (and the top maxima) in $R$ and $T$ appear. In Fig \ref{fig_ss_suboscillations}-b we have plotted $\log_{10}R$ and $\log_{10}T$ for a little larger range of $\alpha$ ($1.98 \leq \alpha \leq 2$) as compared to Fig \ref{fig_ss_suboscillations} -a. Again it is evident from the figures that each sub-peak for $E>E_{ss}^{\alpha=2}$ of non-Hermitian SQM develops a simultaneous maxima in $R$ and $T$ for some values of $\alpha<2$ in non-Hermitian SFQM. This is a new features of scattering in non-Hermitian SFQM and shows that non-Hermitian SFQM is more flexible for spectral singularity. A closer look of the development of SS from the first sub-peak of Fig \ref{fig_ss_suboscillations}-a (also Fig \ref{fig_ss_suboscillations}-b) is shown in Fig \ref{fig_ss_closelook} which shows that the development of SS is also oscillatory in $\alpha$. This is further illustrated in Fig \ref{ss_suboscillation_alpha}.     
\paragraph{}
It is also observed that for $E>E_{ss}^{\alpha=2}$ of non-Hermitian SQM, the increasing order of the energy of sub-peaks  corresponds to the respective decreasing order of $\alpha$ values ($\alpha <2$) in non-Hermitian SFQM. This explains the blue shift of SS energy in non-Hermitian SFQM with reducing $\alpha$. Similarly we have checked that the sub-peaks occurring for  $E<E_{ss}^{\alpha=2}$ in the Fig \ref{fig_ss_suboscillations},  develop SS for $\alpha>2$. However as the range of $\alpha$ is limited to $1 < \alpha \leq 2$ in SFQM, therefore we are not discussing the case of $\alpha >2$.  

The observation mentioned above shows that the sub-peaks around SS in non-Hermitian SQM  have a relation to the development of SS in non-Hermitian SFQM . A mathematical description about this new phenomena would be much interesting and provide a deeper understanding about relation between SS in non-Hermitian SQM and non-Hermitian SFQM.   

\subsection{Coherent perfect absorption in non-Hermitian SFQM}
The condition for coherent perfect absorption (CPA) is 
\begin{equation}
t_{l}t_{r}- r_{l}r_{r}=0
\label{cpa_condition1}
\end{equation}
with $t_{l}=\frac{1}{m_{22}}=t_{r}$ and $r_{l}=\frac{m_{21}}{m_{22}}$, $r_{r}=\frac{m_{12}}{m_{22}}$. In terms of the elements of the transfer matrix, the condition \ref{cpa_condition1} can be written as
\begin{equation}
m_{12} m_{21}=1
\label{cpa_condition2}
\end{equation}
For complex barrier in SFQM this gives the following transcendental equation
\begin{equation}
\mu_{2}^{2} \sin^{2}{\overline{k}_{\alpha}b}=1
\label{cpa_condition_barrier}
\end{equation}
Therefore in order to find the CPA we numerically evaluate the quantity
\begin{equation}
C (E, \alpha)=t_{l}(E, \alpha)t_{r}(E, \alpha)- r_{l}(E, \alpha)r_{r}(E, \alpha)
\label{C_formula}
\end{equation}
and study the variation of $\log{|C|}$ or $\log(\frac{1}{|C|})$ with respect to $E$ and $\alpha$ for a given non-Hermitian barrier potential. We identify CPA as the deep minima in $\log{|C|}$ (or maxima in $\log{(\frac{1}{|C|})}$).

\begin{figure}
\begin{center}
\includegraphics[scale=0.5]{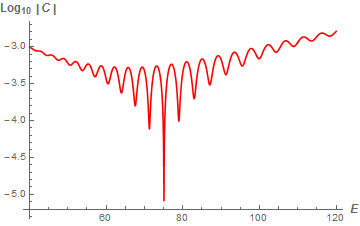} a \includegraphics[scale=0.5]{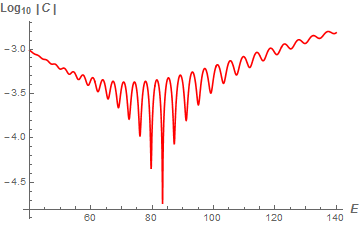} b \\
\includegraphics[scale=0.5]{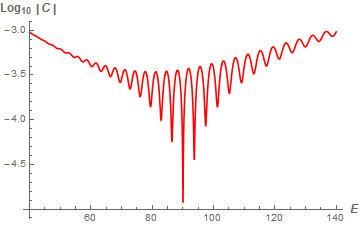} c \includegraphics[scale=0.5]{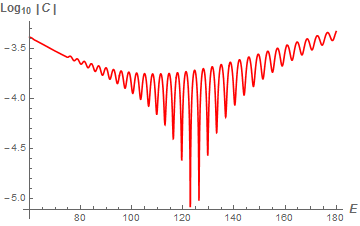} d   
\caption{\it The deep minima in $C(E,\alpha)$ for the potential $V=0.1+5i$ for different values of $\alpha$. Here $b=10$ and $v=10^{-5} $.For the figure $a$, $b$, $c$ and $d$ the values of $\alpha$ are $2$, $1.99$, $1.98$ and $1.95$ respectively. The CPA energy is blue shifted as $\alpha$ decreases.} 
\label{fig_cpa}
\end{center}
\end{figure}
\begin{figure}
\begin{center}
\includegraphics[scale=0.7]{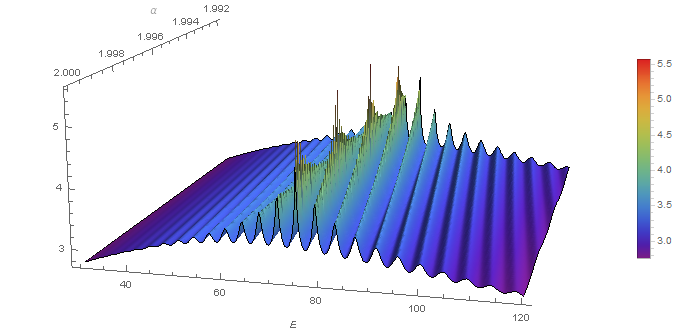}  
\caption{\it Plot showing the development of multiple CPA for $\alpha \leq 2$ over $\alpha-E$ space. The potential parameters are the same as of Fig \ref{fig_cpa} and the range of $\alpha$ is $1.992 \leq \alpha \leq 2$. The color bar is shown adjacent to the figure. It is evident that that the sub-peaks for $E> E_{CPA}^{\alpha=2}$ develop CPA for some values of $\alpha<2$. $E_{CPA}^{\alpha=2}$ is the CPA energy for $\alpha=2$ i.e. non-Hermitian .} 
\label{3d_cpa1}
\end{center}
\end{figure}

\begin{figure}
\begin{center}
\includegraphics[scale=0.5]{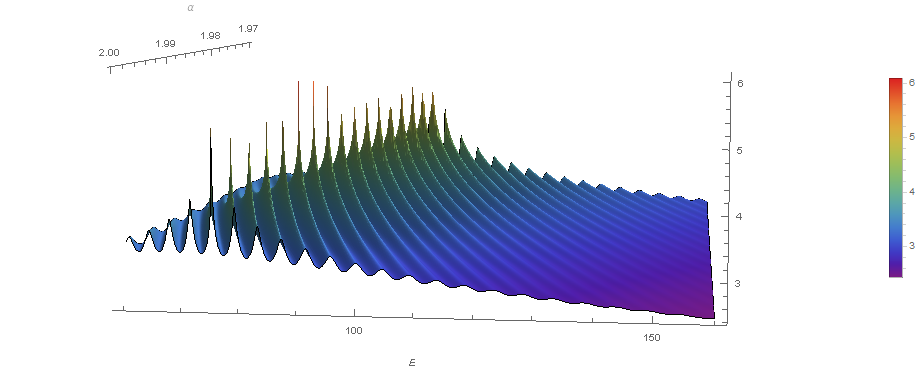}  
\caption{\it Plot showing the development of multiple CPA for $1.97 \leq \alpha \leq 2$ (longer range as compared to Fig \ref{3d_cpa1})over $\alpha-E$ space . The potential parameters are the same as of Fig \ref{fig_cpa} (or Fig \ref{3d_cpa1}). The color bar is shown adjacent to the figure.  There are total $17$ sub-peaks in the figure for the chosen range of energy $E> E_{CPA}^{\alpha=2}$ in the plot.} 
\label{3d_cpa2}
\end{center}
\end{figure}

\begin{figure}
\begin{center}
\includegraphics[scale=0.5]{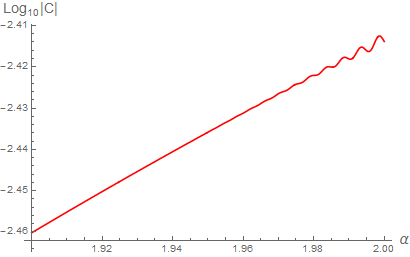} a \includegraphics[scale=0.5]{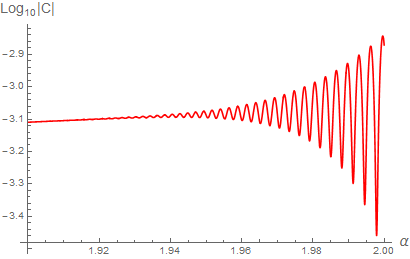} b \\
\includegraphics[scale=0.5]{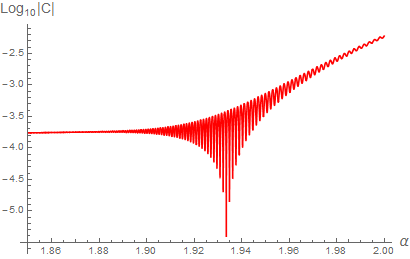} c \includegraphics[scale=0.5]{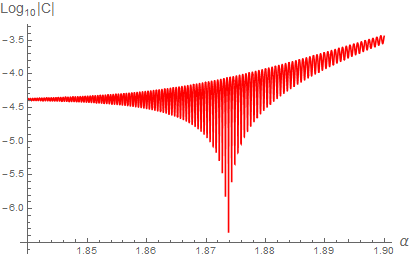} d \\
\includegraphics[scale=0.5]{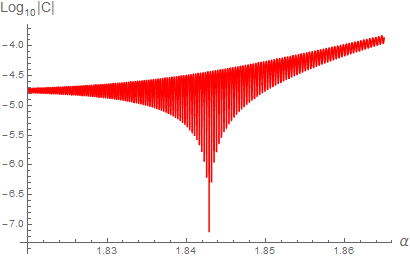} e \includegraphics[scale=0.5]{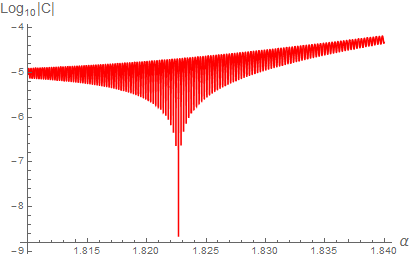} f \\
\includegraphics[scale=0.55]{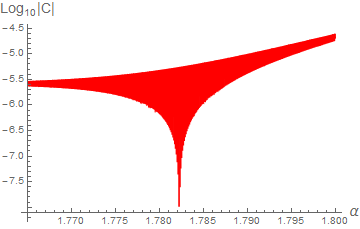} g \includegraphics[scale=0.55]{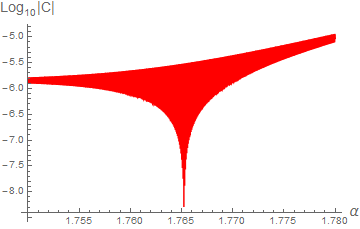} h \\    
\caption{\it Plots showing the variation of $\log_{10}C$ with $\alpha$ at different energies $E$ for the same potential parameter as of Fig \ref{fig_cpa}. For Fig-a,b,c,...,h the respective energies are $E=50, 100, 200, 400, 600, 800, 1500, 2000$. This again illustrate that for higher energy, the CPA corresponds to comparatively lower $\alpha$ values. It is also observed that CPA occurring at low $\alpha$ values (and higher energies) have more dense sub-peaks with $\alpha$ in the vicinity of CPA. }  
\label{fig_cpa_alpha}
\end{center}
\end{figure}

Fig \ref{fig_cpa} shows the plot of $\log{(|C|)}$-$E$ for different value of $\alpha$ for the potential $V=0.1+5i$ with $b=10$. It is seen that deep minima of $C(E,\alpha)$ is blue shifted with decreasing $\alpha$.  To understand the development of CPA, we $3$D plot $\log{(\frac{1}{|C|})}$ with $E$ and $\alpha$. The $3$D plot is shown in Fig \ref{3d_cpa1} for the same potential parameters as in Fig \ref{fig_cpa}. We see that for standard case of $\alpha=2$, $\log{\frac{1}{C}}$ curve have sub-peaks (similar to the case of SS) with energy $E$ on both side of the maxima. It is observed that the sub-peaks for $\alpha=2$ that occur for energy $E> E_{\alpha=2}^{CPA}$ develop maxima for some value of $\alpha<2$. Here $E_{\alpha=2}^{CPA}$ is the CPA energy for non-Hermitian SQM ($\alpha=2$). Similar to the case of SS, it is also observed that for $E>E_{CPA}^{\alpha=2}$ of non-Hermitian SQM, the increasing order of the energy of sub-peaks of $\log{\frac{1}{C}}$ with $E$  corresponds to the respective decreasing order of $\alpha$ values ($\alpha <2$) in non-Hermitian SFQM. This also explains the blue shift of CPA energy with decreasing $\alpha$. We have also observed that the sub-peaks for $E<E_{\alpha=2}^{CPA}$ develop CPA for some values of $\alpha>2$. However we are not considering the case of $\alpha>2$ as the range of $\alpha$ in SFQM is limited to $1<\alpha \leq 2$. Fig \ref{3d_cpa2} shows the development of CPA for $1.97 \leq \alpha \leq 2$. We also observe that similar to sub-peaks with $E$ around the CPA, there are sub-oscillations with $\alpha$ in the neighbourhood of CPA (see Fig \ref{3d_cpa1}). The occurrence of these sub-oscillations with $\alpha$ in the neighbourhood  of CPA is further graphically illustrated in Fig \ref{fig_cpa_alpha} at different energies as sub-minima. It is noticed that the CPA with lesser $\alpha$ values have more dense sub-minima with $\alpha$ around CPA.   

The CPA features mentioned above are similar to the features of SS discussed earlier. This further indicates that the sub-minima of CPA in non-Hermitian SQM (when $\alpha=2$) case has a relation to the development of CPA in non-Hermitian SFQM (when $\alpha <2$). Again a mathematical description of these new CPA features would be required to provide a deeper understanding about relation between the CPA in non-Hermitian SFQM and non-Hermitian SQM.              

\section{Conclusions and Discussions}
\label{the_result}
The scattering features of non-Hermitian quantum mechanics, SS and CPA are investigated in the domain of non-Hermitian SFQM with Levy index $1< \alpha \leq 2$ for non-Hermitian delta and rectangular barrier potentials. The case of non-Hermitian delta potential is analytically studied. It is found that for delta potential $V(x)=-i\rho \delta (x-x_{0})$, $\rho \in R^{+}$, the SS energy is blue shifted with decreasing $\alpha$ if $2 \hbar v < \rho$ where $v$ is the characteristic velocity of the non-relativistic system. In this case as $\alpha \rightarrow 1$ the energy points of spectral singularity move to infinity. However for potential strength such that $4 \hbar v > \rho e$ the energy points of spectral singularity are red shifted with decreasing $\alpha$ and move to zero as $\alpha \rightarrow 1$. For potential strength such that $2 \hbar v=\rho$ we have $ E_{ss}^{\alpha =2} \leq E_{ss}^{\alpha} < \frac{e}{2} E_{ss}^{\alpha=2}$ for   $1 < \alpha \leq 2$.   

The case of non-Hermitian rectangular barrier is investigated numerically. It is found that non-Hermitian SFQM has more flexibility for SS and CPA to occur at different energies by varying $\alpha$. The most notable and interesting feature is due to the observation that the subsequent sub-peaks in SS/CPA of standard NHQM for energy $E> E_{SS/CPA}^{\alpha=2}$ develop SS/CPA  for values of $\alpha <2$ in subsequent decreasing order. This observation shows that the sub-peaks around SS/CPA in non-Hermitian SQM (when $\alpha=2$) case has a relation to the development of SS/CPA in non-Hermitian SFQM (when $\alpha <2$). A mathematical description about this new phenomena would be much interesting and will provide a deeper understanding about relation between SS and CPA in non-Hermitian SQM and non-Hermitian SFQM.

{\it \bf{Acknowledgements}}: \\
MH acknowledges support from SAG/ISAC and encouragement from Dr. M. Annadurai, Director, ISAC to carry out this research work. BPM acknowledges the support from CAS, Department of Physics, BHU.

\end{document}